# A Semi-Supervised Inf-Net Framework for CT-Based Lung Nodule Analysis with a Conceptual Extension Toward Genomic Integration

Fateme Mobini[a], Mohammad Reza Hedyehzadeh[a], Mahdi Yousefi[a]

[a] Department of Biomedical Engineering, Azad University, Dezful, Khuzestan, Iran

## Abstract

Lung cancer is a primary contributor to cancer-related mortality globally, highlighting the necessity for precise early detection of pulmonary nodules through low-dose CT (LDCT) imaging. Deep learning methods have improved nodule detection and classification; however, their performance is frequently limited by the availability of annotated data and variability among imaging centers. This research presents a CT-driven, semi-supervised framework utilizing the Inf-Net architecture to enhance lung nodule analysis with minimal annotation. The model incorporates multi-scale feature aggregation, Reverse Attention refinement, and pseudo-labeling to efficiently utilize unlabeled CT slices. Experiments conducted on subsets of the LUNA16 dataset indicate that the supervised Inf-Net attains a score of 0.825 on 10,000 labeled slices. In contrast, the semi-supervised variant achieves a score of 0.784 on 20,000 slices that include both labeled and pseudo-labeled data, thus surpassing its supervised baseline of 0.755. This study presents a conceptual framework for the integration of genomic biomarkers with CT-derived features, facilitating the development of future multimodal, biologically informed CAD systems. The proposed semi-supervised Inf-Net framework improves CT-based lung nodule assessment and lays the groundwork for flexible multi-omics diagnostic models.

## Introduction

A significant percentage of cancer-related deaths globally are caused by lung cancer, which is still one of the deadliest cancers [4]. Though many instances are still discovered at advanced stages, when treatment options are limited and the prognosis is poor, early detection greatly increases the five-year survival chance [4]. This difficulty highlights the need for more trustworthy and physiologically based diagnostic techniques. When compared to chest radiography, low-dose computed tomography (LDCT) has shown significant mortality reductions and has become the standard imaging method for early lung cancer screening [3], [5], [6]. However, LDCT produces a lot of pictures and has a significant false-positive rate, especially when it comes to small, low-contrast, or anatomically complex nodules like juxta-pleural and juxta-vascular lesions [6], [26]. Consistent CT interpretation is made more difficult by inter-observer variability among radiologists, which frequently results in needless follow-up procedures and uncertainty when separating benign from malignant nodules [3], [5]. By automating processes including detection, segmentation, and malignancy prediction, computer-aided diagnostic (CAD) systems have been created to lessen these constraints [1], [31]. Due to scanner heterogeneity and limited representational capacity, traditional methods based on manually created radiomic features, intensity descriptors, and texture metrics have demonstrated promise, but their performance deteriorates in heterogeneous, multi-center CT datasets [8], [15], and [18]. By directly learning hierarchical characteristics from imaging data, deep learning-based models—in particular, convolutional neural networks, or CNNs—have greatly improved lung nodule analysis [7], [30]. Large, well-annotated datasets are necessary for these models, but obtaining them is challenging because of time limits, privacy concerns, and annotation costs. Patients with similar radiological profiles—the same nodule morphology, stage, and grade—may have notably divergent clinical outcomes, according to clinical evidence that goes beyond imaging alone. This disparity emphasizes the essential part that molecular and genetic factors play in determining tumor behavior. Tumor aggressiveness, responsiveness to treatment, and survival are significantly impacted by changes in genes including EGFR, KRAS, TP53, RB1, and BCL2, as well as by changes in gene expression and epigenetic modifications [1], [4], [12], [23], [28], and [37]. These biological elements offer information that cannot be obtained from CT scans alone. These findings encourage the integration of genetic and radiological data as a means of achieving more accurate and customized diagnosis. Molecular markers may provide an explanation for why two patients who appear identical on CT may have very distinct clinical trajectories. This idea is fundamental to the rationale for this research and is backed by a wealth of biological and oncological evidence. However, matched genetic biomarkers are absent from popular CT datasets like LUNA16 and LIDC-IDRI, which hinders a thorough experimental assessment of multimodal fusion techniques [12], [23], and [28]. When only CT scans are available, CAD frameworks must (1) function reliably and (2) be architecturally ready for genomic integration when such datasets become available. This paper creates a semi-supervised framework for CT-driven lung nodule identification and classification based on the Inf-Net architecture in order to meet these demands. Effective learning of delicate pulmonary structures under sparse annotation is made possible by Inf-Net's multi-scale feature aggregation, Parallel Partial Decoder (PPD), Reverse Attention (RA) processes, and Edge Attention modules [29], [40], and [41]. By using unlabeled CT slices, a pseudo-labeling technique further minimizes the need for huge, labeled datasets [17], [29]. Building on this CT-based foundation, we present a conceptual extension that describes how high-level CT representations may be fused with genomic descriptors in future multimodal applications, including methylation patterns, expression signatures, and mutation indicators [12], [23], [28], and [37]. This offers a foundation for next-generation diagnostic systems that is motivated by biology. This research is innovative in that it develops a CT-based, semi-supervised model that is specifically made to support future genomic integration while framing a genomically informed diagnostic perspective. When matching datasets become available, our twofold contribution guarantees (1) robust performance on existing CT datasets and (2) a clear path toward biologically enriched, multi-omics CAD systems. This paper's remaining sections are arranged as follows: The epidemiology of lung cancer, genetic risk factors, imaging-based detection techniques, and CAD frameworks are reviewed in Section 2. The suggested methodology, which includes the Inf-Net architecture, semi-supervised learning approach, and conceptual CT–genomic fusion module, is presented in Section 3. The experimental setup and outcomes are described in detail in Section 4. Future directions are concluded at the end of Section 5 and Section 6.

## Background and Related Work

Lung cancer accounts for a significant percentage of worldwide cancer mortality, with early detection being crucial for enhancing survival rates [4], [25]. Low dose computed tomography (LDCT) effectively identifies early pulmonary nodules; however, the interpretation of the substantial volume of CT slices produced during screening poses challenges for radiologists [3], [5], [24]. The variability in nodule size, shape, density, and anatomical location complicates visual assessment, especially for juxta-pleural and juxta-vascular nodules that closely resemble adjacent structures [6], [26]. Figure 1 presents typical examples of solitary, juxta-pleural, and juxta-vascular nodules frequently observed in LDCT scans, which often pose challenges for both automated and manual interpretation.

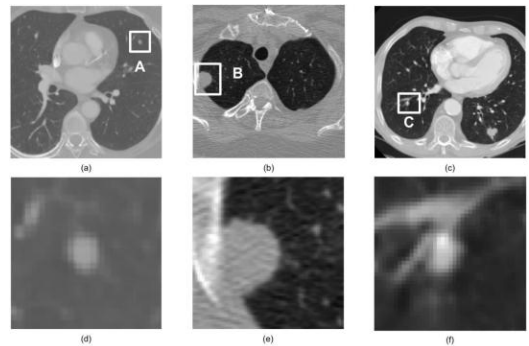

**Figure 1.** Three main types of pulmonary nodules in CT scans: (a) solitary nodule, (b) juxta-pleural nodule attached to the chest wall, and (c) juxta-vascular nodule located near vessels. These categories differ in spatial location and segmentation difficulty, with juxta-pleural and juxta-vascular nodules posing significant challenges for CAD systems

Traditional computer-aided detection (CAD) systems for lung nodules have developed over several decades and generally consist of four primary components: lung segmentation, candidate nodule detection, feature extraction, and benign/malignant classification [27]. Previous approaches predominantly utilized handcrafted features, including size, shape descriptors, texture metrics, and curvature measures [2], [3], [5], [9], [11], [14], [17], [18–22]. Although these methods demonstrated satisfactory performance on specific datasets, their dependence on manually crafted features constrained their generalizability across different scanners, slice thicknesses, and patient demographics [8], [15–18], [20]. Deep learning has significantly transformed the analysis of lung nodules. Convolutional neural networks (CNNs) have shown significant efficacy in nodule detection, segmentation, and malignancy prediction by acquiring hierarchical and task-specific representations directly from imaging data [7], [10], [12], [13], [16], [19], [30], [32], [33]. Multi-view CNNs, 3D CNNs, and hybrid feature-learning models demonstrate efficacy in segmenting small and subtle nodules, while also minimizing false positives [9], [34], [35], [39]. These models generally necessitate significant quantities of labeled data, which are challenging to obtain in clinical settings due to the considerable costs and labor involved in expert annotation. In addition to imaging, substantial evidence associates lung tumor behavior with genetic and molecular changes, including oncogene activation and tumor-suppressor inactivation [1], [4], [12], [23], [28], [37]. Existing CT datasets, including LIDC-IDRI and LUNA16, lack matched genomic biomarkers. However, the integration of imaging and genomic descriptors represents a significant research focus, potentially enhancing malignancy prediction by elucidating biological processes not discernible in CT scans. Semi-supervised learning approaches have garnered significant interest as they facilitate effective model training with limited annotation. Methods utilizing pseudo-labeling and self-training enable models to exploit extensive amounts of unlabeled CT slices, thereby reducing the need for annotations [17], [29], [41]. Inf-Net, initially developed for the segmentation of COVID-19 infections, employs multi-scale feature aggregation via a Parallel Partial Decoder (PPD) and enhances segmentation accuracy through Reverse Attention (RA) modules, showcasing its effectiveness in delineating intricate pulmonary patterns [29], [41]. In summary, previous research identifies three primary challenges: (1) the limited availability of fully annotated CT datasets, (2) the necessity for effective feature extraction under diverse imaging conditions, and (3) the increasing significance of multimodal integration to include biological context. The identified challenges drive the creation of a semi-supervised framework based on Inf-Net, which optimizes the use of unlabeled CT data and establishes a basis for future genomic-radiologic integration.

## 3. Methodology

### 3.1 Overview

This research presents a CT-driven semi-supervised framework utilizing the Inf-Net architecture to enhance lung nodule detection and classification in scenarios with limited annotations. Inf-Net, initially developed for COVID-19 CT infection segmentation [29], incorporates multi-scale feature aggregation and refinement mechanisms, rendering it effective for the identification of subtle pulmonary structures. Due to the scarcity and high cost of annotated CT data, the framework employs a pseudo-labeling strategy influenced by previous semi-supervised methods in medical image segmentation [17], [29], [41]. Furthermore, while LUNA16 lacks matched genomic biomarkers, a conceptual framework is presented that outlines how future multimodal datasets may incorporate genetic descriptors, informed by biological insights connecting genomic alterations to tumor aggressiveness. [12], [23], [28], [37].

### 3.2 Baseline Models

To contextualize the performance of the proposed Inf-Net-based framework, two baseline approaches were implemented: a convolutional neural network (CNN) segmentation model and a semi-supervised support vector machine (SVM). The CNN baseline employs a standard encoder–decoder architecture frequently utilized in lung lesion segmentation [34], [39]. Although these models effectively capture coarse spatial structures, their performance is constrained by the inability to differentiate visually ambiguous nodule boundaries. This aligns with findings that purely morphological cues are frequently inadequate for reliable malignancy prediction [30]. The second baseline consists of a semi-supervised SVM classifier that utilizes handcrafted radiomic features, which encompass intensity statistics, texture descriptors, and shape metrics, in accordance with established CAD methodologies [12], [36], [38]. Despite optimization efforts, the SVM baseline exhibited diminished robustness, primarily attributable to the constrained expressiveness of handcrafted descriptors and their susceptibility to scanner variability—issues extensively documented in previous research [3], [4], [6], [15–18]. The identified shortcomings underscore the necessity for enhanced feature extractors that can effectively capture multi-scale and contextual information directly from CT scans, thereby justifying the selection of Inf-Net as the primary backbone for this study.

### 3.3 Inf-Net Architecture for CT Feature Extraction

Inf-Net is chosen as the backbone because it effectively integrates global multi-scale representations with local refinement mechanisms [29], [41]. This capability is crucial for analyzing lung nodules, which exhibit significant variability in size, density, and shape, and are frequently associated with the pleura or vasculature. The architecture comprises three interrelated components: The Parallel Partial Decoder (PPD) consolidates high-level features derived from the encoder's deepest layers, generating a coarse global guidance map that identifies potential nodule regions [29]. In contrast to classical decoders that depend significantly on low-level features, PPD selectively integrates only high-level feature maps to minimize computational complexity while maintaining robust semantic signals. This enhances the localization of nodules across different scales.

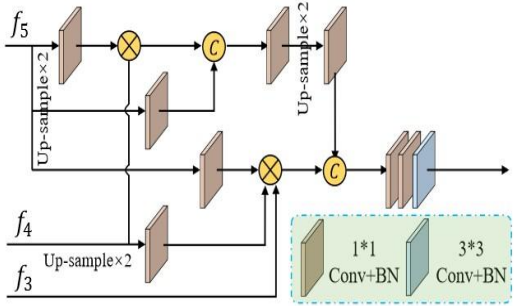

**Figure 2.** Parallel Partial Decoder (PPD) used to aggregate multi-scale high-level features into a global guidance map

Inf-Net utilizes Reverse Attention (RA) modules to iteratively enhance segmentation accuracy in uncertain or low-contrast regions by refining the coarse PPD prediction [41]. RA functions by inverting the current prediction map to pinpoint uncertain areas, enabling the model to systematically rectify false positives and false negatives. This method is especially efficacious for nodules located near arteries or pleural margins, where intensity congruence may deceive conventional segmentation models.

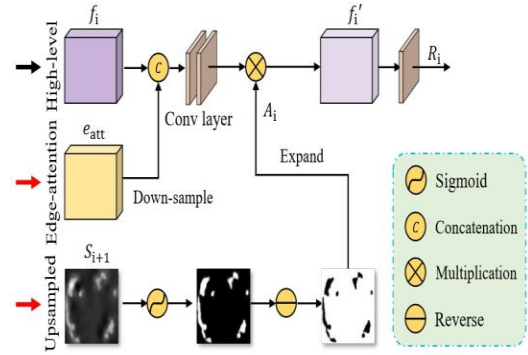

**Figure 3.** Reverse Attention (RA) module is used to refine predictions by focusing on hard regions guided by the global map and edge features.

The Edge Attention (EA) mechanism enhances boundary precision by utilizing shallow feature maps to identify edge structures [40], [41]. EA enhances supervision through edge maps, aiding the network in delineating fine-grained boundaries and enhancing segmentation consistency for small or irregular nodules. The complete Inf-Net architecture incorporates PPD, RA, and EA modules, employing deep supervision across intermediate outputs [29]. This design facilitates concurrent learning of global structure, local refinement, and boundary precision, establishing a robust feature extraction framework for subsequent classification tasks.

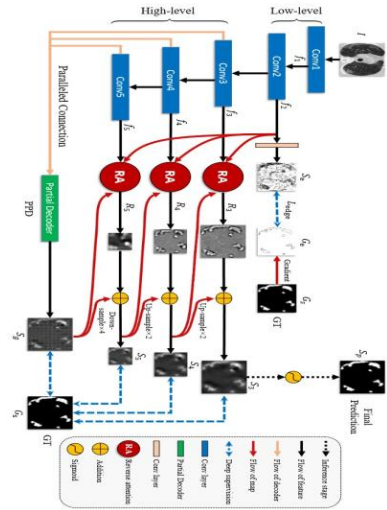

**Figure 4.** The Inf-Net architecture consists of three Reverse Attention (RA) modules connected to a Parallel Partial Decoder (PPD)

### 3.4 Semi-Supervised Learning with Pseudo-Labeling

A semi-supervised pseudo-labeling strategy is utilized to enhance the training data, given the restricted size of the annotated subset of LUNA16. The approach adheres to self-training principles that are frequently applied in semi-supervised medical segmentation [17], [29], [41]. An initial Inf-Net model is trained using the labeled slices that are available. This model is subsequently employed to produce pseudo-labels for unlabeled CT slices. Predictions that surpass a set confidence threshold are kept, minimizing confirmation bias and avoiding the buildup of errors. With every iteration, additional pseudo-labeled slices are incorporated into the training set, and the model undergoes fine-tuning utilizing both ground-truth and high-confidence pseudo-labels. This process, through several iterations, gradually enhances the diversity of the training data, allowing the model to encounter nodules with different shapes, contrast levels, and anatomical contexts.

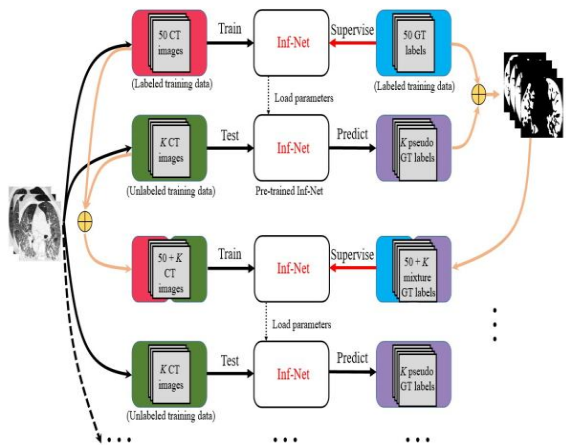

**Figure 5.** Semi-supervised training framework of the proposed Inf-Net–based model with pseudo-label generation and iterative dataset expansion

Particularly well-suited to large-scale CT datasets, this method improves Inf-generalization Net's performance without necessitating substantial human annotation.

### 3.5 Conceptual Extension for CT–Genomic Fusion

The LUNA16 dataset utilized in this study lacks matched genomic information; however, the integration of genetic biomarkers with CT-derived features is a significant avenue for the advancement of next-generation CAD systems [12], [23], [28], [37]. Genetic alterations, including mutations in EGFR, KRAS, TP53, RB1, and BCL2, as well as gene-expression and epigenetic signatures, significantly impact tumor aggressiveness, treatment response, and survival [1], [4], [12], [23], [28], [37]. Molecular factors offer additional biological insights that are not entirely represented by CT morphology alone. Integrating genomic descriptors with radiological representations offers a promising approach for enhancing the precision and individualization of lung nodule characterization. We propose a conceptual multimodal fusion pipeline to prepare for future datasets that integrate imaging and genomic profiles, extending the existing CT-driven model into a biologically enriched diagnostic framework. In this design, Inf-Net functions as the main feature extractor for CT slices, producing multi-scale and high-level semantic representations that capture nodule morphology, boundary characteristics, and contextual anatomical cues [29], [41]. Concurrently, genomic descriptors—such as mutation indicators, mRNA expression signatures, or methylation-derived biomarkers—will undergo preprocessing steps that include normalization, feature selection, and dimensionality reduction (e.g., PCA or curated gene panels) [12], [23], [28]. The integration of multimodal data will take place via an intermediate fusion module, in which genomic feature vectors are combined with the deep CT features generated by Inf-Net. This joint representation would subsequently be provided to a downstream classifier, such as fully connected layers, which can learn complex interactions between morphological and molecular cues [24]. This design allows the model to integrate complementary modalities. CT imaging offers structural insights into nodule characteristics, whereas genomic biomarkers provide biological context concerning tumor progression, mutation-driven behavior, and malignancy risk [12], [23], [28], [37]. This multimodal integration is presented as a conceptual extension rather than an experimentally evaluated component of the current study. The absence of publicly accessible CT datasets with associated genomic labels necessitates that the proposed fusion framework acts as a foundational model for subsequent research and system advancement. Incorporating genomic descriptors is anticipated to improve the accuracy of malignancy predictions, address inter-patient variability not captured by imaging data alone, and facilitate more personalized and biologically informed decision-making in early lung cancer screening [12], [23], [28], [37]. This multimodal extension is presented conceptually and was not assessed due to limitations in the dataset.

### 3.6 Loss Functions and Optimization

In order to get the best possible optimization, we have taken a page out of previous research on edge-aware salient object recognition and infection segmentation [29], [41]:

- Edge Loss ($L_{edge}$): A binary cross-entropy (BCE) loss calculated between the predicted edge map and the ground-truth edge map obtained via hand annotations.

$$L_{edge} = -\sum_{x=1}^{w}\sum_{y=1}^{h}\left[G_{e(x,y)}\log(S_e) + \left(1 - G_{e(x,y)}\right)\log(1 - S_e)\right] \quad (3.1)$$

- Segmentation Loss ($L_{seg}$): A weighted aggregation of Intersection-over-Union (IoU) loss and weighted Binary Cross-Entropy (BCE), which prioritizes "hard" pixels adjacent to lesion boundaries and in uncertain areas.

$$L_{seg} = L_{wIoU} + \lambda L_{wBCE} \quad (3.2)$$

- Total Loss ($L_{total}$): When training is stabilized by deep supervision, this unified loss is computed across the global map (PPD output) and all side outputs from RA modules [29].

$$L_{total} = L_{seg}(G_s, S_g^{up}) + L_{edge} + \sum_{i=3} L_{seg}(G_s, S_i^{up}) \quad (3.3)$$

The evaluation of model performance employs segmentation metrics, specifically the Dice score and IoU, alongside classification metrics such as accuracy, precision, and recall, in alignment with previous studies on lung nodule analysis [30], [32], [33].

## 4. Experiments and Results

This section outlines the implementation settings, dataset preparation, training procedures, and evaluation strategy employed to assess the performance of the Inf-Net and Semi-Inf-Net models. All experiments were conducted using Python with the PyTorch framework. Training was performed on an NVIDIA GPU, such as the RTX series, equipped with 16 to 32 GB of system memory. The Inf-Net backbone adheres to the reference implementation outlined in [29], incorporating architectural refinements and supervisory signals specified in [40] and [41]. The Adam optimizer was employed with a learning rate of 1e-4, $β_1$ set to 0.9, $β_2$ set to 0.999, and a batch size of 8.

### 4.1 Backbone Network Evaluation

We evaluated ResNet50, Res2Net50, and ResNet101 to find the best backbone network for Inf-Net. To give an accurate picture, we trained all the models using the identical preprocessing, training, and assessment parameters. In line with previous research, overfitting and instability on tiny labelled subsets meant that deeper backbones did not always increase performance.

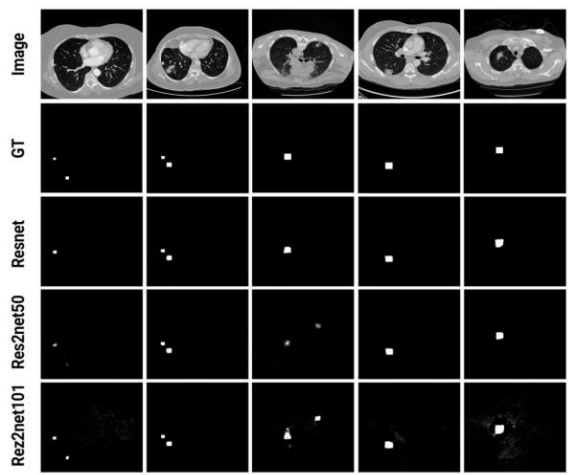

Figure 6. Comparison of different backbone networks used in Inf-Net, including ResNet50, Res2Net50, and ResNet101. ResNet50 provides the best balance between accuracy and stability.

All future tests used ResNet50 as their default backbone, based on these results.

### 4.2 Dataset Preparation

Experiments were conducted using the LUNA16 dataset, which offers annotated CT volumes for the analysis of pulmonary nodules. Each scan underwent preprocessing that included resampling to achieve uniform voxel spacing, clipping of intensities to a specified Hounsfield Unit (HU) range, and normalization to minimize inter-scan variability. Lung masks were created to exclude non-essential structures, including the trachea and mediastinum, retaining only slices that contained nodules or adjacent regions to concentrate training on anatomically significant areas for diagnosis. The semi-supervised learning framework outlined in Section 3 involved partitioning the dataset into two components: a labeled subset for training the initial supervised Inf-Net model and a significantly larger unlabeled subset for pseudo-label generation and iterative dataset expansion.

### 4.3 Training Protocol

The training occurred in two consecutive phases to facilitate the semi-supervised learning framework. Initially, a supervised Inf-Net model was trained just on the labeled subset of LUNA16 to develop preliminary segmentation and feature extraction capabilities. The model was subsequently employed in the second phase to provide pseudo-labels for the unlabeled CT slices, retaining high-confidence predictions in accordance with methodologies established in previous semi-supervised medical image segmentation studies [17], [29], [41]. The freshly pseudo-labeled data were amalgamated with the original labeled samples, and the model was refined on this augmented dataset with the combined loss outlined in Section 3, which incorporates weighted BCE, IoU loss, edge-aware BCE, and deep supervision. During both phases, the Adam optimizer was employed with uniform hyperparameters, and early halting was implemented based on validation performance to avert overfitting and guarantee steady convergence.

### 4.4 Performance Evaluation

The model's performance was assessed on the reconstructed test subset of the LUNA16 dataset employing uniform preprocessing, post-processing, and slice-level assessment criteria across all methodologies. The supervised Inf-Net model with a ResNet50 backbone functioned as the principal reference, with its accuracy detailed in Table 1. These findings establish a robust baseline performance standard for CT-based nodule detection utilizing high-quality labeled data.

| Test Set | Number of Images | Training Time | Accuracy |
|---|---|---|---|
| Test 1 | 60 | 32 min | 0.536 |
| Test 2 | 4000 | 11 h | 0.792 |
| Test 3 | 10000 | 56 h | 0.825 |

Table 1. Inf-Net accuracy for lung nodule detection

Table 1 illustrates that the supervised Inf-Net setup demonstrates robust initial performance; nevertheless, its learning capacity is constrained by the relatively small size of the labeled training sample. The Semi-Inf-Net system addresses this limitation by utilizing pseudo-labeled slices derived from the extensive unlabeled segment of the dataset. Table 2 delineates the accuracy of the Semi-Inf-Net model and underscores its enhancement relative to the supervised version.

| Test Set | Number of Images | Labeling Time | Training Time | Accuracy |
|---|---|---|---|---|
| Test 1 | 10000 | 271 h | 56 h | 0.755 |
| Test 2 | 20000 | 648 h | 114 h | 0.784 |

Table 2. Semi-Inf-Net accuracy for lung nodule detection

The juxtaposition of Table 1 and Table 2 illustrates the advantages of semi-supervised learning. Semi-Inf-Net enhances generalization by integrating high-confidence pseudo-labels, thereby encountering a wider variety of nodule sizes, shapes, and situations. This enhancement is also apparent qualitatively. Figure 7 illustrates visual comparisons of outputs from Inf-Net and Semi-Inf-Net, demonstrating that the semi-supervised model yields more coherent borders and more comprehensive detections, particularly for low-contrast and juxta-vascular nodules.

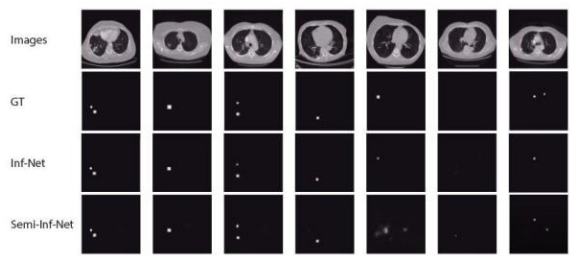

Figure 7 — Inf-Net vs Semi-Inf-Net Output Comparison

Table 3 presents a comprehensive comparison of the performance of the CNN baseline, semi-supervised SVM, supervised Inf-Net, and Semi-Inf-Net across all implemented algorithms. The findings indicate that conventional handcrafted-feature methods (CNN and SVM) are inferior to deep learning-based frameworks, with Semi-Inf-Net attaining the highest overall accuracy among all assessed models, hence validating the efficacy of the semi-supervised approach.

| Rank | Team | Score |
|---|---|---|
| 1 | PAtech (PA_tech) | 0.951 |
| 2 | JianPeiCAD (weiyixie) | 0.95 |
| 3 | LUNA16FONOVACAD (zxp774747) | 0.947 |
| 4 | iFLYTEK-MIG (yinbaocai) | 0.941 |

| 5 | zhongliu_xie (zhongliu.xie) | 0.922 |
|---|---|---|
| 6 | iDST-VC (chenjx1005) | 0.897 |
| 7 | qfpxfd (qfpxfd) | 0.891 |
| 8 | CASED (CASED) | 0.887 |
| 9 | 3DCNN_NDET (lishaxue3) | 0.882 |
| 10 | Aidence (mjharte) | 0.871 |
| 11 | junxuan20170516 (chenjx1005) | 0.865 |
| 12 | MEDICAI (bharadwaj) | 0.862 |
| 13 | Ethan20161221 (ethanhwang2012) | 0.856 |
| 14 | resnet (QiDou) | 0.839 |
| 15 | CCELargeCubeCnn (Intel_wuhui) | 0.833 |
| 16 | ZNET (gzuidhof) | 0.811 |
| 17 | MOT_M5Lv1 (elopez69) | 0.742 |
| 18 | VisiaCTLung (jacobsc) | 0.715 |
| 19 | etrocad (jefvdmb2) | 0.676 |
| 20 | M5LCADThreshold0.3 (atraverso) | 0.608 |
| 21 | lance (lancesy) | 0.543 |
|  | **Our Inf-Net (10,000 Images)** | **0.825** |
|  | **Our Semi-Inf-Net (20,000 Images)** | **0.784** |

Table 3. comparison of suggested methods

The evaluation indicates that the proposed semi-supervised Inf-Net framework consistently enhances performance compared to supervised models and significantly surpasses traditional feature-based methods. Both quantitative and qualitative findings substantiate the significance of integrating unlabeled data and utilizing Inf-Net's multi-scale refinement modules for enhanced nodule detection.

## 5. Conclusion and Future Work

### 5.1 Summary of Findings

This research introduced a CT-driven, semi-supervised framework utilizing the Inf-Net architecture for the detection and classification of pulmonary nodules in scenarios with limited annotations. Preliminary experiments employing traditional image-based methods, including a semantic segmentation convolutional neural network and a semi-supervised support vector machine, demonstrated restricted discriminative ability, particularly for ambiguous, low-contrast, or anatomically intricate nodules. The results highlighted the necessity for more advanced feature extractors that can directly capture multi-scale morphological cues from CT imagery. Inf-Net was modified to enhance nodule-specific feature extraction through multi-scale context aggregation, Reverse Attention refinement, and edge-aware supervision to overcome these limitations. The framework was enhanced through the integration of a pseudo-labeling strategy that utilized the extensive collection of unlabeled CT slices in the LUNA16 dataset. Experimental evaluation revealed that the supervised Inf-Net attained a score of 0.825 on 10,000 labeled slices, whereas the semi-supervised variant achieved 0.784 when trained on 20,000 combined labeled and pseudo-labeled slices, indicating an enhancement over its supervised baseline of 0.755. The findings indicate that the integration of Inf-Net with a semi-supervised learning strategy improves robustness and generalization in scenarios with limited annotated data.

This study highlighted the importance of integrating genomic information into the early diagnosis of lung cancer, in addition to CT-based modeling. Genetic alterations, such as mutations in EGFR, KRAS, TP53, RB1, and BCL2, significantly influence tumor aggressiveness, treatment response, and individual patient variability, providing additional molecular insights beyond what CT morphology can reveal. A conceptual multimodal fusion framework was proposed to demonstrate the integration of genomic descriptors with Inf-Net features in future datasets, despite LUNA16's lack of matched genomic data. This outlines a method for advancing next-generation CAD systems that integrate morphological and molecular biomarkers to enhance the accuracy of malignancy predictions, contingent upon the availability of suitable imaging-genomic datasets.

### 5.2 Conclusion

This study's results indicate that the semi-supervised Inf-Net framework offers a robust and effective solution for CT-based lung nodule analysis, especially in contexts with limited labeled data. The integration of high-confidence pseudo-labels enables the model to decrease dependence on extensive manual annotation while preserving robust predictive performance. The supervised and semi-supervised variants demonstrated consistent enhancements compared to traditional image-based baselines, validating the efficacy of Inf-Net's multi-scale feature extraction and refinement mechanisms for intricate pulmonary structures. This study establishes a clear architectural foundation for the future integration of molecular biomarkers, despite the inability to evaluate full multimodal learning due to the lack of genomic annotations in the LUNA16 dataset. This paper presents a conceptual CT–genomic fusion strategy that demonstrates the integration of radiological and genomic descriptors to tackle inter-patient biological variability not accounted for by CT morphology alone. The proposed framework provides a scalable and extensible foundation for the development of advanced CAD systems that can adapt to clinical variability and facilitate earlier, biologically informed lung cancer diagnosis.

### 5.3 Limitations

Several limitations must be recognized in the interpretation of these findings. Initially, all experiments were limited to CT-only analysis due to the absence of matched genomic or multi-omics profiles in commonly utilized public datasets, such as LUNA16. The proposed multimodal CT–genomic fusion strategy could not be empirically evaluated. The semi-supervised learning process employed a limited subset of unlabeled slices. Expanding pseudo-labeling to encompass the entire dataset or implementing more sophisticated confidence-filtering techniques could potentially enhance performance further. The present evaluation relies on slice-level predictions, which may inadequately reflect volumetric continuity or patient-level heterogeneity. The integration of 3D architectures or hybrid 2.5D models may enhance the understanding of nodules that extend across multiple slices. Fourth, the performance of Inf-Net may be affected by domain shifts due to variations in scanner hardware, reconstruction protocols, or institutional imaging practices. This underscores the necessity for multi-center validation to confirm robustness and generalizability in practical clinical settings.

### 5.4 Recommendations for Future Work

Future research may expand this framework in several significant areas. Initially, assessing the proposed CT–genomic fusion strategy through datasets that encompass both imaging and genomic profiles would facilitate a formal evaluation of multimodal performance and enhance personalized malignancy prediction. Secondly, investigating alternative semi-supervised or self-supervised techniques—such as consistency regularization, teacher–student architectures, contrastive pretraining, or diffusion-based pseudo-label refinement—could enhance model robustness in scenarios with limited annotation. Third, the integration of 3D volumetric processing, hybrid 2.5D methods, or transformer-based encoders could improve representation learning for nodules that extend across multiple slices or display subtle spatial patterns. Fourth, exploring contemporary backbone networks like EfficientNet, DenseNet, Swin Transformer, or other lightweight architectures may enhance both efficiency and accuracy. Large-scale, multi-center clinical evaluations across various imaging systems, institutions, and patient populations are crucial for addressing domain shifts and demonstrating real-world reliability.